# Data-driven Discovery of 3D and 2D Thermoelectric Materials

Kamal Choudhary, Kevin F. Garrity and Francesca Tavazza

Materials Science and Engineering Division, National Institute of Standards and Technology, Gaithersburg, Maryland 20899, USA.

**ABSTRACT**

In this work, we first perform a systematic search for high-efficiency three-dimensional (3D) and two-dimensional (2D) thermoelectric materials by combining semiclassical transport techniques with density functional theory (DFT) calculations and then train machine-learning models on the thermoelectric data. Out of 36000 three-dimensional and 900 two-dimensional materials currently in the publicly available JARVIS-DFT database, we identify 2932 3D and 148 2D promising thermoelectric materials using a multi-steps screening procedure, where specific thresholds are chosen for key quantities like bandgaps, Seebeck coefficients and power factors. We compute the Seebeck coefficients for all the materials currently in the database and validate our calculations by comparing our results, for a subset of materials, to experimental and existing computational datasets. We also investigate the effect of chemical, structural, crystallographic and dimensionality trends on thermoelectric performance. We predict several classes of efficient 3D and 2D materials such as $Ba(MgX)_2$ (X=P,As,Bi), $X_2YZ_6$ (X=K,Rb, Y=Pd,Pt, Z=Cl,Br), $K_2PtX_2$(X=S,Se), $NbCu_3X_4$ (X=S,Se,Te), $Sr_2XYO_6$ (X=Ta, Zn, Y=Ga, Mo), $TaCu_3X_4$ (X=S, Se,Te), and XYN (X=Ti, Zr, Y=Cl, Br). Finally, as high-throughput DFT is computationally expensive, we train machine learning models using gradient boosting decision trees (GBDT) and classical force-field inspired descriptors (CFID) for n-and p-type Seebeck coefficients and power factors, to quickly pre-screen materials for guiding the next set of DFT calculations. The dataset and tools are made publicly available at the websites: https://www.ctcms.nist.gov/~knc6/JVASP.html , https://www.ctcms.nist.gov/jarvisml/ and https://jarvis.nist.gov/ .

**Corresponding author:** Kamal Choudhary (E-mail: kamal.choudhary@nist.gov)

# I INTRODUCTION

Thermoelectrics[1-4] are materials that can convert a temperature gradient into electric voltage, or vice-versa. Themoelectrics can be used to regenerate electricity from waste heat[5], refrigeration[6] and several other space-technology applications[7,8]. The search for efficient thermoelectric materials is an area of intense research due the potential of converting waste heat into electrical power, and therefore improving energy efficiency and reducing fossil fuel usage. The figure of merit for thermoelectric materials is the dimensionless quantity $zT$:

$$zT = \frac{S^2\sigma}{k_e+k_l}T \qquad (1)$$

where $S$, $\sigma$, $k_e$, $k_l$, and $T$ are the Seebeck coefficient, electrical conductivity, electronic part of thermal conductivity, lattice thermal conductivity, and temperature, respectively. The numerator, $S^2\sigma$, is referred to as the power-factor. To achieve a high $zT$, a material should have a high-power factor and low thermal conductivity. Experimental synthesis and characterization are ultimately the critical steps to prove the usefulness of a thermoelectric material; however, experiments are costly and time-consuming, and the list of potential thermoelectrics is very large. Computational methods based on first principles density functional theory (DFT) can be very useful in the initial screening process, as well as in interpreting experimental results. DFT[9-12] has successfully predicted the Seebeck coefficients and power factors for various classes of bulk materials. There has also been a series of high-throughput computational searches for bulk/three dimensional (3D) thermoelectrics. Chen et al.[9] and Ricci et al.[13] compute thermoelectric properties of more than 48000 materials and show a reasonably strong comparison between the maximum Seebeck-coefficient determined from DFT and experiment for a subset. Garrity[14,15] use high-throughput first-principles calculations to screen transition metal oxides, nitrides, and sulfides for candidate materials with high power factors and low thermal conductivity. Gorai et al.[16] develop TEDesignLab as a thermoelectrics-focused virtual laboratory that contains calculated thermoelectric properties using several thermoelectric metrics[15]. Carrete et al. use high-throughput method to identify materials with low-thermal conductivity[17]. He et al.[18] search for several transition metal oxides with high thermoelectric performance. In addition to the computational databases, there are several developments of experimental databases also such as UCSB-MRL thermoelectric database[19]. Similar to the 3D materials, there has been a huge upsurge in research on monolayer/two-dimensional (2D) materials due to their promising high Seebeck coefficients and low thermal conductivities[20-24]. Despite the above research, a systematic, combined database that allows for the comparison of bulk and monolayer thermoelectric properties is still lacking. In addition, such a systematic database of thermoelectric properties is necessary to develop machine learning models for predicting the thermoelectric properties of new materials, which would circumvent the high computational cost of additional DFT calculations and potentially guide materials discovery. There have been a few recent reports on the applications of machine learning for thermoelectric properties[25-27], but the field is still developing.

In this work, first, we present a high-throughput DFT database of bulk and monolayer thermoelectric properties. In this high-throughput work, we focus on finding high power factor materials, which is a necessary and less computationally expensive step in identifying thermoelectrics. We do not attempt to predict zT values in this work. All of the data and tools are provided at the JARVIS-DFT website, which is a part of the materials genome initiative (MGI) at the National Institute of Standards and Technology (NIST). The JARVIS-DFT database contains about 36000 bulk and 900 low-dimensional materials with their DFT-computed structural, exfoliability[28], elastic[29], optoelectronic[30] solar-cell efficiency[31], and topologically non-trivial[32] properties. Using this database, we highlight a few novel 3D-bulk materials and 2D-monolayer materials that we predict have good thermoelectric properties. We also investigate correlations of thermoelectric properties with chemistry and structure of materials. Finally, we develop highly accurate machine learning models for quickly identifying efficient thermoelectric materials.

**II METHODS**

The DFT calculations are carried out using Vienna Ab-initio simulation package (VASP)[33,34] software using the JARVIS-DFT workflow given on our github page (https://github.com/usnistgov/jarvis ). Please note commercial software is identified to specify procedures. Such identification does not imply recommendation by NIST. We use OptB88vdW functional[35], which gives accurate lattice parameters for both vdW and non-vdW (3D-bulk) solids[28]. There have been several tests of vdW functionals[28,63] and OptB88vdW predicts important physical quantities such as lattice constants, bulk modulus and atomization energies as well as or better than other available vdW functionals. We employ spin-polarized calculations, starting with a ferromagnetic spin-ordering, during the geometric optimization of each material. In this work, a material is termed as low-dimensional if it contains vdW-bonding in one (2D-bulk), two (1D-bulk), and three (0D-bulk) crystallographic directions. Details of the low-dimensional material database can be found in Ref.[28,29]. A monolayer/2D-material is simulated with broken periodicity in z-direction with a vacuum padding of at least 18 Å. The transport properties were calculated using the Boltzmann transport equation (BTE) implemented in the BoltzTrap code[37]. The BTE is used to investigate the non-equilibrium behavior of electrons and holes by statistically averaging all possible quantum states given by the equation below:

$$\frac{df(\boldsymbol{k},T,t)}{dt} = \left(\frac{\partial f(\boldsymbol{k},T,t)}{\partial t}\right)_s - \frac{d\boldsymbol{k}}{dt}\nabla_k f(\boldsymbol{k},T,t) - v(\boldsymbol{k})\nabla_r f(\boldsymbol{k},T,t). \qquad (2)$$

$f$ is the electron distribution, which is a function of state $k$, temperature $T$ and time $t$, and $v(k)$ are the electron group velocities. The three terms on the right-hand side of Eq. 2 refer, respectively, to the temporal rate of change of $f$ due to all scattering processes, the rate of change of $f$ due to external forces, and the diffusion from the carrier density gradient. If the external forces consist only of a low electric field, E, and no magnetic field, B, such that $\frac{dk}{dt} = \frac{eE}{h}$ then the low-filed BTE is given by:

$$\frac{df(\boldsymbol{k},T,t)}{dt} + v(\boldsymbol{k})\nabla_r f(\boldsymbol{k},T) + \frac{eE}{h}\nabla_k f(\boldsymbol{k},T) = \left(\frac{\partial f(\boldsymbol{k},T,t)}{\partial t}\right)_s. \tag{3}$$

Now, $f$ can be described as a first-order (linear) perturbation from the (equilibrium) Fermi-Dirac distribution, $f_0$, due to scattering

$$\left(\frac{\partial f(\boldsymbol{k},T,t)}{\partial t}\right)_s = -\frac{f(k)-f_0(k)}{\tau} \tag{4}$$

where

$$f_0[\mathcal{E}(k)] = \frac{1}{e^{[\mathcal{E}(k)-\mathcal{E}_F]/k_B T}+1} \tag{5}$$

where the dependence of $\mathcal{E}$ on $k$ is given by the electronic band structure, and the various scattering terms and time dependence are lumped into the electronic relaxation time, $\tau$. The computation of the relaxation time is very computationally expensive, especially in a high-throughput context[38]. The BoltzTrap code uses constant relaxation time approximation (CRTA) and the rigid band-approximation (RBA). In CRTA, the relaxation time cancels out for Seebeck coefficients, but for electrical conductivity, we choose a value of $10^{-14}$ $s$ as the relaxation time[39]. The relaxation time can depend on both intrinsic factors like electron-phonon coupling and extrinsic factors like the presence of defects. The RBA assumes that the shape of density of states does not change by doping or increasing the operation temperature. This methodology has been used earlier for both bulk[20-24] and monolayer[40] materials. We converge the k-points and plane wave cut-off in DFT-calculations for all the materials in our database using energy convergence criteria[41] of 0.001 eV. These k-points and cut-offs values are generally higher than the usual 1000/atom k-points selected in previous high-throughput database studies. Note that for some metallic and low-bandgap materials it is difficult to converge the BoltzTrap calculations, hence their transport values are not reported. Calculation of thermal conductivity is very computationally expensive because they require the calculation of anharmonic force constants. So, we limit such calculations to only one example material. The force constants are fit with the method described in Garrity[14, 15] and BTE is solved using ShengBTE[42].

The machine-learning models are trained using classical force-field inspired descriptors (CFID) descriptors[43] and supervise machine learning techniques using decision-trees (DT), random-forest (RF), k-nearest neighbor (KNN), multilayer perceptron (MLP), gradient boosting techniques in the scikit-learn package[44] also GBDT implemented in XGBoost[45] and LightGBM[46] packages. The CFID gives a unique representation of a material using structural (such as radial, angle and dihedral distributions), chemical, and charge descriptors. Using the DFT thermoelectric data, we train classification machine learning models to identify whether a material has Seebeck-coefficient less than -100 µV/K for n-type and more than 100 µV/K for p-type, n and p-power factor more than 1000 µW/(mK)$^2$ at 600 K and $10^{20}$/cm$^3$ doping. The CFID[43] has been recently used to develop several high-accuracy ML models for material properties such as formation energies, bandgaps, refractive index, bulk and shear modulus and exfoliation energies k-points, cut-offs[41], and solar-cell efficiencies[31].

## III RESULTS AND DISCUSSION

We use BoltzTrap to calculate the electronic transport properties for all the 36000 bulk and 900 monolayer materials in our database, calculating the Seebeck coefficients (S), electrical-conductivities ($\sigma$), power factors, and electronic part of thermal conductivities, all as a function of temperature and doping. The resultant data is used for screening potential thermoelectric materials, analyzing trends and machine learning training processes. A flow-chart describing our computational search is shown in Fig. 1.

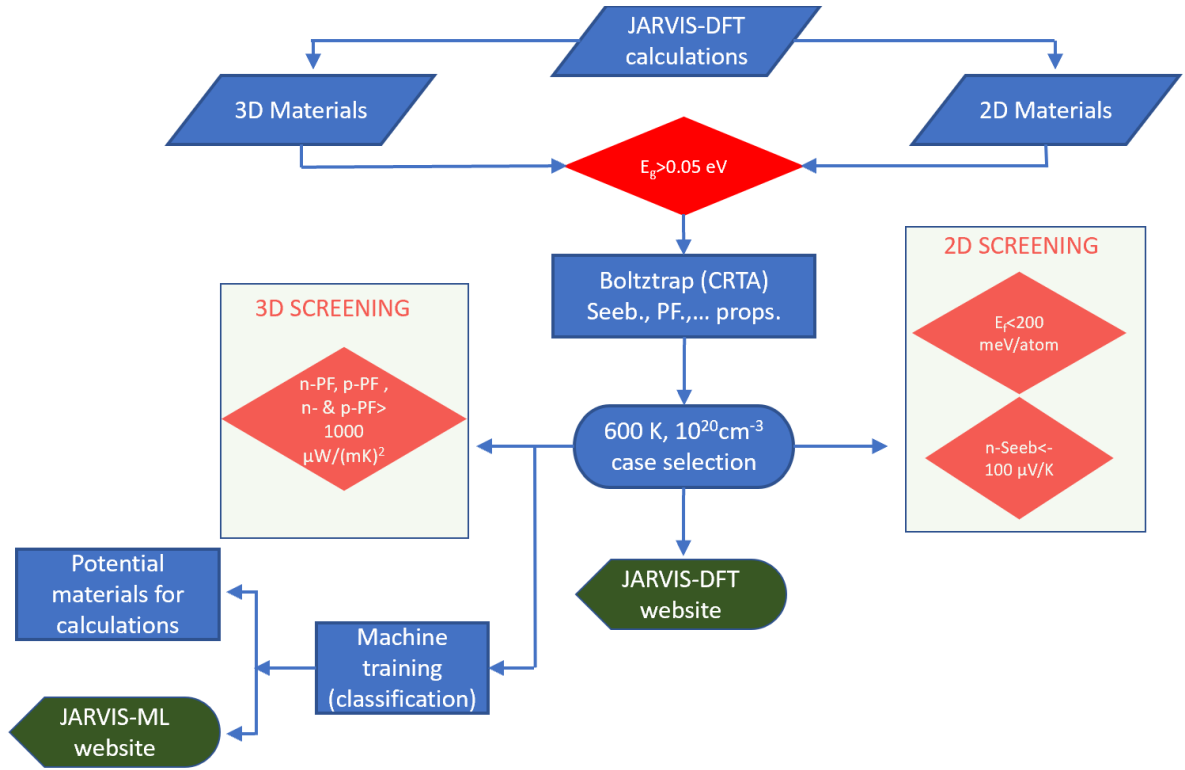

*Fig. 1 Flow chart associated with the data-driven thermoelectric materials design.*

### IIIA EXPERIMENTAL AND THEORETICAL BENCHMARKING

To benchmark our computational methodology, we compare our Seebeck coefficients for a subset of 14 materials to experimental data, and we find a mean absolute deviation (MAD) of 54.7 µV/K ($r^2$=0.94). The details of this comparison are shown in the supplementary information (Fig. S1a and Table. S1). Next, we compare our Seebeck coefficients (JV) to DFT results from another computational database, the Materials-Project (MP)[13], to ascertain how sensitive our results are to specific choices of DFT parameters (exchange-correlation functional, k-points density and energy cutoff). We look at 9434 compounds from the MP and JV databases, comparing the n-type Seebeck coefficient at 600K and $10^{20}$/cm$^3$ doping. We obtained a MAD of 18.8 µV/K ($r^2$=0.87), signifying that DFT data for Seebeck coefficients are closer to each other than to the experimental results. We attribute the differences between the MP and JV datasets to the fact that MP uses the GGA-PBE[47] functional as well as fixed k-points and cutoffs for their DFT calculations, while JV uses the OptB88vdW functional and an automatic convergence procedure for k-points and cutoffs, which we expect to provide improved results especially for vdW-bonded materials. More details/data on the DFT comparisons are also provided in the supplementary information (Fig. S1b).

**IIIB SCREENING AND ANALYSIS OF BULK THERMOELECTRICS**

After computing the Seebeck coefficients for all the materials in the database, we concentrate on developing a screening method that identifies high-efficiency thermoelectric materials. The thermoelectric performance of a material depends on several quantities, most notably, temperature, doping type, and doping concentration. We chose to concentrate on temperatures of 600 K and $10^{20}$ /cm$^3$ doping, which represent a typical thermoelectric operating temperature and a doping level that is achievable for many semiconductors. However, we note that dopability of a semiconductor depends on several critical factors such as native defect energetics, which are too computationally expensive to predict from first principles in a systematic manner. Many experimentally relevant thermoelectrics are doped at similar concentration as shown in the supplementary information (Table. S1). As good thermoelectric materials are generally semiconducting or insulating, we first screen for materials with bandgap>0.05 and BoltzTrap data, which narrows the search set from 36000 down to 8764 materials. We present an overview of the database in Fig. 2. The n and p-type Seebeck coefficients are generally negative and positive values, respectively, with a maximum absolute value of 600 µV/K, as shown in Fig. 2a and 2d. In Fig. 2b and 2e, we show the distribution of power factors for n-type and p-type materials. Experimentally known high-efficiency thermoelectric have power factors of more than 1000 µW/(mK)$^2$. Although a high Seebeck-coefficient is necessary for a high power factor, there is typically a significant tradeoff between S and $\sigma$, necessitating a more careful analysis[14]. This tradeoff can be seen in Fig. 2c, which shows a scatter plot of S versus $\sigma$ for n-type thermoelectrics. The sizes of dots are proportional to the bandgaps and color-coded based on their power factor values. A similar inverse-relationship applies to p-type materials as well. In Fig 2f, we look at the relationship between power factors and band gaps, finding that high power factors occur more often in low band gap materials. A similar relationship holds for the n-type materials as well.

For the next step in the screening procedure, we select only materials with n-type and p-type power factor at 600 K and $10^{20}$/cm$^3$ larger than 1000 µW/(mK)$^2$, which gives us 4330 and 4403 candidates, respectively. We find 2932 materials with both n-and p-type PF >1000 µW/(mK)$^2$. For this case, we analyzed the set of selected materials in terms of various chemical and physical attributes. To begin with, we classified their dimensionality, which is determined by lattice-constant and data-mining approaches[28]. As shown in Fig. 2g, we find that 14.4 % of the high efficiency thermoelectrics are low-dimensional, i.e. vdW-bonded, while the rest have three-dimensional bonding. Next, we find that these materials tend to be highly symmetric, as cubic and trigonal symmetry materials are over-represented (as shown in Fig. 2h). As discussed earlier, we find a range of Seebeck coefficients among our set of high-power factor materials, due to the tradeoff between S and $\sigma$, as shown in Fig. 2i. Finally, we note that our screening process rediscovered several well-known thermoelectrics, such as $Bi_2Te_3$, SnSe, GeTe, $Mg_2Si$, $PtSe_2$, PbSe, PbTe and $PtSe_2$, as well as many potential new thermoelectrics. Some examples of families of thermoelectrics uncovered in our screening include $Ba(MgX)_2$ (X=P,As,Bi), $X_2YZ_6$ (X=K,Rb, Y=Pd,Pt, Z=Cl,Br), $K_2PtX_2$(X=S,Se), $NbCu_3X_4$(X=S,Se,Te), $Sr_2XYO_6$(X=Ta, Zn, Y=Ga, Mo), $TaCu_3X_4$(X=S, Se,Te). Some of these materials are 0D-bulk i.e. with vdW interactions along all three axes, as in $AsI_3$ (JVASP-3636), or along just two axes (1D-bulk materials such as BiSeI (JVASP-5200), $TeBr_2$ (JVASP-33839), $SeI_2$ (JVASP-33798)), or, as in most cases, along one axis only (2D-bulk materials such as BN (JVASP-17), $YSnF_5$ (JVASP-8344), $HfS_2$ (JVASP-210), $MnBr_2$ (JVASP-2041), GeTe (JVASP-1157), $TiO_2$ (JVASP-30586), $NiO_2$ (JVASP-8645), GaP (JVASP-28372), $SbAsO_4$ (JVASP-10177)). Many other newly identified materials for thermoelectrics applications are 3D-bulk materials, i.e. no vdW-bonding, such as LiMgN (JVASP-22546), $CaO_2$ (JVASP-22677 ), $Li_2CuSb$ (JVASP-7820), $SrZrO_3$ (JVASP-8037), $K_2PtSe_2$ (JVASP-2838), $Al_2ZnS_4$ (JVASP-9688), $ZnO_2$ (JVASP-10252), $Sr_2TaGaO_6$ (JVASP-10974), $TiSnO_3$ (JVASP-35817), ZrSiPt (JVASP-40824), $YAlO_3$ (JVASP-50410), AlAs (JVASP-8183), $RbAuC_2$ (JVASP-7652). A full list is provided in the supplementary information. We will provide information on the stability of each compound using a convex-hull approach on the website soon. The JVASP-# denotes the JARVIS-IDs of the materials the details of which can be found at corresponding https://www.ctcms.nist.gov/~knc6/jsmol/JVASP-#.html. For example, the details of JVASP-8037 is available at https://www.ctcms.nist.gov/~knc6/jsmol/JVASP-8037.html.

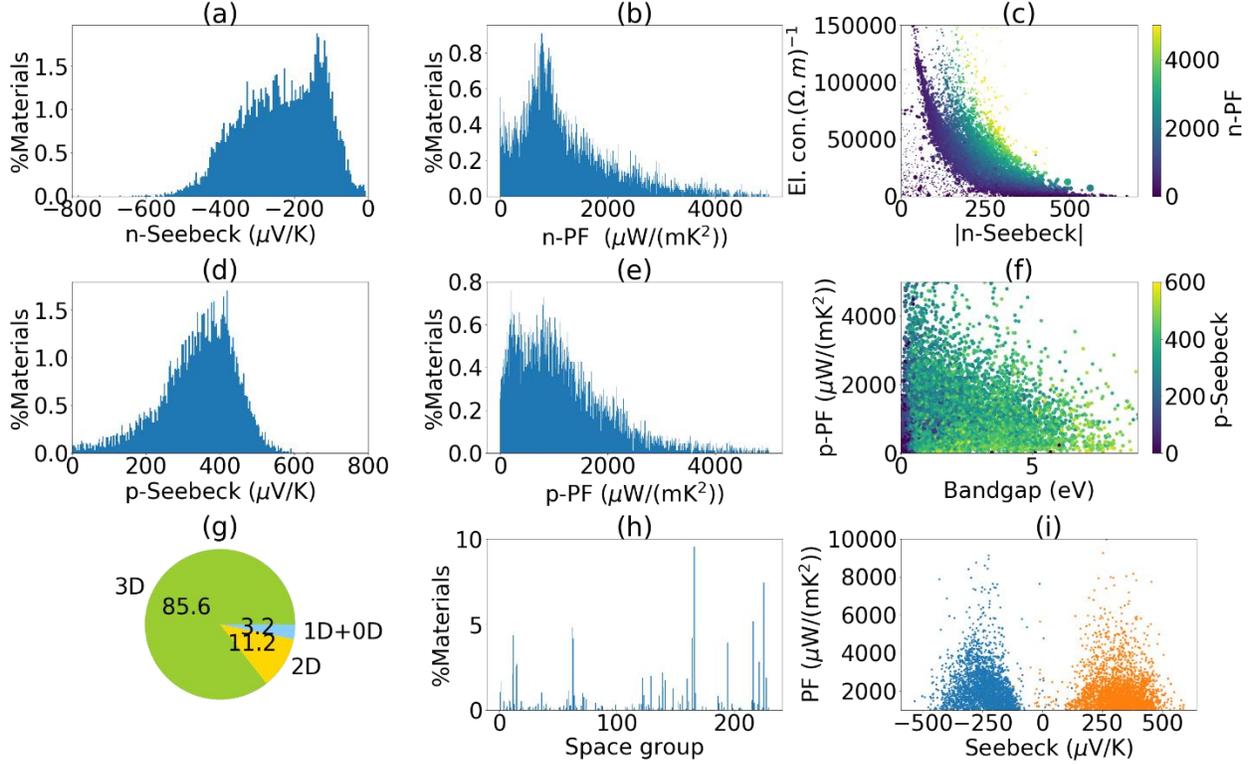

*Fig. 2 A brief overview of the thermoelectric data for periodic bulk materials. Figure a) -f) have been computed on all 3D materials, while g-i) display properties only computed on the set of bulk screened materials (bandgap>0.05 eV, n-type and p-type PF>1000 (µW/(mK)$^2$) at 600 K and $10^{20}$ cm$^{-3}$ doping concentration). a) n-type Seebeck coefficient distribution, b) n-type power factor of materials, c) n-type electrical conducitivity plotted against the absolute values of Seebeck-coefficient with color-coded power-factor and size of the dots proportional to bandgaps, d) p-type Seebeck coefficient distribution, e) p-type power factor of materials, f) p-type power factor plotted against the bandgaps, g) predicted dimensionality distribution of screened materials, h) space-group distribution of the screened materials, i) power-factor vs Seebeck distribution of the screened materials.*

In Fig. 3, we show the likelihood that a compound containing a given element has a high-power factor. More specifically, for every compound containing a given element, we calculate the percentage-probability that those materials have an n-power-factor greater than 1000 µW/(mK)$^2$. We find that many of the alkaline earth metals, early transition metals, Ir, Pt, Cu, Ag and chalcogenides were found to contribute towards high efficiency materials, which is again consistent with previously known thermoelectric materials[1-4] such as $Bi_2Te_3$, SnSe, GeTe, $Mg_2Si$, $PtSe_2$. For example, 298 out of 587 Se-containing compounds in our database have power-factors greater than 1000 $\mu W/(mK)^2$ so, the percentage is 50.77%. Such periodic table trends can help guide new materials searches or doping strategies that may result in improved thermoelectrics. As there are no clear trends between high-PF materials and common chemical characteristics, such as the electronegativity, that would be obvious on a periodic table, there is a need for more sophisticated statistical machine learning techniques, which we carry out in a following section.

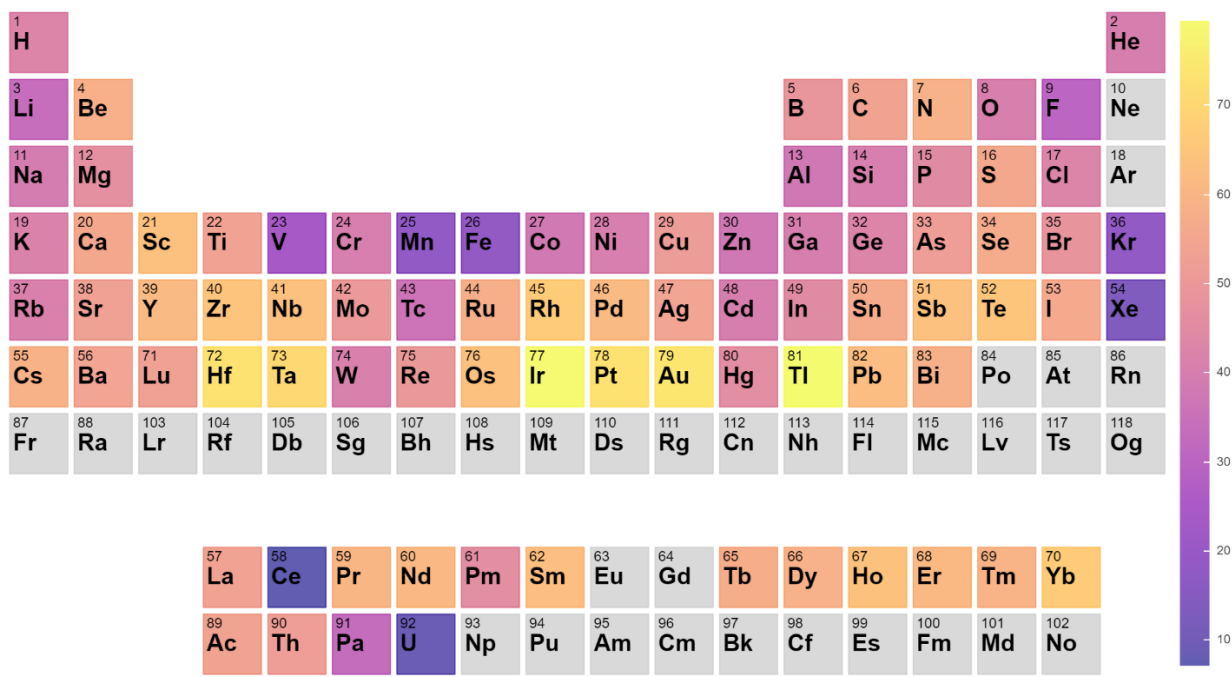

*Fig. 3 Periodic table trends of high-power factor materials. The elements in a material were assigned 1 or 0 if the material has high or low n-type power-factor (>1000 ($\mu W/(mK)^2$)). Then the probability of finding the element in a high power-factor material was calculated.*

**IIIC SCREENING AND ANALYSIS OF MONOLAYER THERMOELECTRICS**

The screening process described and analyzed up to this point was applied exclusively to bulk materials. Next, we apply a similar procedure to monolayer 2D materials. The vdW-bonded materials (2D-bulk) can be exfoliated to produce 2D-monolayers, which have shown promise in thermoelectric applications[20-24]. The exfoliability of a 2D-bulk material is depends on its exfoliation energy, as described in detail in Ref.[28]. In order to find exfoliable high-efficiency 2D monolayers, we select materials in our database which meet the following criteria: a) exfoliation energy < 200 meV/atom, b) bandgap>0.05 and c) monolayer-n-type Seebeck-coefficient <-100 µV/K. In order to compare the thermoelectric properties of monolayer and bulk structures, it is necessary to express the doping and conductivity quantities in a way that does not depend on the arbitrary vacuum thickness of a monolayer simulation cell. To achieve this, we rescaled the volume of the monolayers using the thickness of a 2D-layer. Using this screening procedure, we identified 148 promising 2D monolayers among 900 materials. As seen in Fig. 4a, comparing 2D and 3D Seebeck coefficients, we see that monolayers tend to have smaller absolute Seebeck coefficients than their bulk counterparts. This shows that interlayer coupling is important to thermoelectric behavior, and that 2D thermoelectric properties can't be exactly obtained from their bulk counterpart. Nevertheless, the Spearmen's correlation between the bulk and monolayer Seebeck coefficients is 0.711 while the Pearson's correlation is 0.721, as shown in Table 1 and S2, suggesting noteworthy correlation. We also investigated the correlation between 3D and monolayer density of states (DOS) at the Fermi level, for 600 K and $10^{20}$/cm$^3$ doping, as well as the correlation for the effective mass. Both results are given in Table 1 and S2. We find a strong correlation between bulk and monolayer effective masses (0.80), and an even stronger correlation (0.90) between bulk and monolayer DOS at the Fermi level, suggesting that the Seebeck-coefficient is more difficult to predict. A lower Spearman's correlation of 0.43 was reported for bulk systems Seebeck-coefficient and DOS by Garrity et al. [14] for transition metal oxides, sulfides and nitrides. Kumar et al. [21] also showed similar correlations for monolayer and bulk $WSe_2$, using DFT calculations.

*Table.1 Spearman correlation of monolayer (Mono) and bulk density of states (DOS) (states/unitcell) at Fermi level, the effective mass of electrons ($m_e$) and Seebeck coefficients (μV/K).*

| Spearman correlation | Mono-DOS | Bulk-DOS | Bulk-Seebeck | Mono-Seebeck | Bulk-$m_e$ | Mono-$m_e$ |
|---|---|---|---|---|---|---|
| **Mono-DOS** | - | 0.897 | 0.592 | 0.553 | -0.474 | -0.432 |
| **Bulk-DOS** | 0.897 | - | 0.604 | 0.566 | -0.493 | -0.412 |
| **Bulk-Seebeck** | 0.592 | 0.604 | - | 0.711 | -0.859 | -0.723 |
| **Mono-Seebeck** | 0.533 | 0.566 | 0.711 | - | -0.596 | -0.668 |
| **Bulk-$m_e$** | -0.474 | -0.493 | -0.859 | -0.596 | - | 0.801 |
| **Mono-$m_e$** | -0.432 | -0.412 | -0.723 | -0.668 | 0.801 | - |

Some of the high-efficiency 2D-monolayer materials that we find are: AuBr (JVASP-27756), SnSe (JVASP-5929), SnS (JVASP-19989), $PtSe_2$ (JVASP-744), $ZrS_3$ (JVASP-792), GaSe (JVASP-687), $WSe_2$ (JVASP-652). The full database is available online. We note a special class of vdW-bonded materials, XYZ (X=Ti,Zr, Y=N, Z=Cl,Br,I) (such as JVASP-6268, JVASP-6181, JVASP-6184), which show consistently high thermoelectric behavior, and which would be a promising target for future investigation. While many of these compounds have not yet been experimentally verified as 2D materials[28], they are almost all related to experimentally known 3D layered structures from the ICSD, and have exfoliation energies consistent with the small number of experimentally studied 2D materials. We hope that works like this one will in part encourage the study of a wider range of 2D materials.

As a first step in this direction, we carried out thermal conductivity calculations for a representative example, ZrBrN (JVASP-12027), in its 3D form, which resulted in a remarkable low lattice thermal conductivity of 0.3 W / m K at 300K. The unit cell and phonon dispersion curve of ZrBrN are shown in Fig. 4b and 4c. We see that while the structure is dynamically stable (all positive frequency modes), there are low-frequency phonon modes which extend across the Brillioun zone, which are responsible for the strongly anharmonic behavior in this system. Such detailed analysis for candidate thermoelectric materials will be considered in the future.

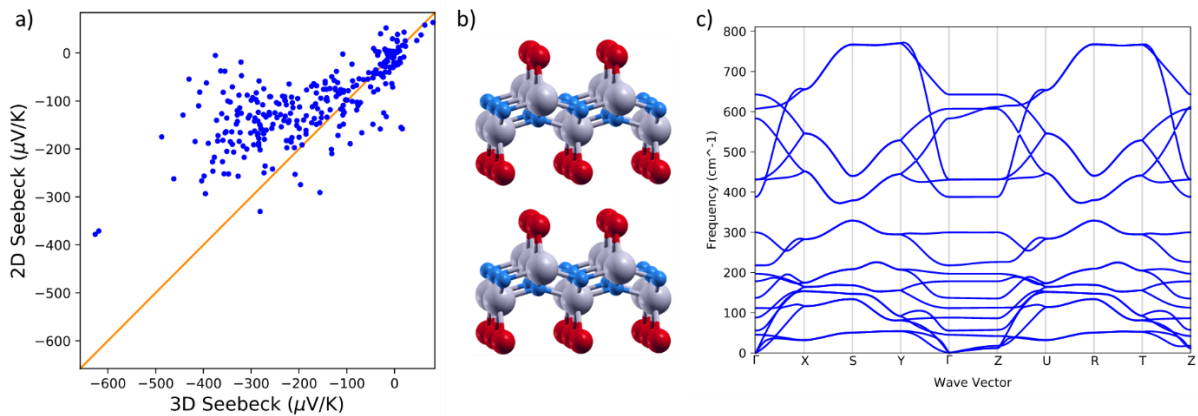

*Fig. 4 a) 3D vs 2D Seebeck coefficients, b) crystal structure of ZrBrN (JVASP-12027), Zr in grey, N in blue, Br in red c) Phonon bandstructure for ZrBrN.*

**IIID MACHINE LEARNING ANALYSIS**

Finally, to accelerate the DFT screening process, we train supervised classification machine learning models[43] for n and p-type Seebeck coefficients and power-factors. In this scheme, we simply classify whether materials have a property value greater or lesser than certain thresholds. The accuracies of the classification models are evaluated based on the area under curve (AUC) of the receiver operating characteristics (ROC) curves. The ROC curve illustrates the model's ability to differentiate between high and low-performance materials, classifying a material to be high-performance if its Seebeck coefficient is less than -100 µV/K for n-type, or more than 100 µV/K for p-type, and if its n- and p-power factor is more than 1000 µW/(mK)$^2$ at 600 K and $10^{20}$/cm$^3$ doping. The ROC curve plots the prediction rate for high-performance materials, correctly versus incorrectly predicted. A value of 0.5 implies random guessing, while a value of 1.0 implies a perfect model. We first train classification models with default parameters using decision tress (DT), random forest (RF), k-nearest neighbors (KNN), multi-layer perceptron (MLP), and gradient boosting models implemented in scikit-learn package and also GBDT implemented in XGBoost (XGB) and LightGBM (LGB) packages. As a standard practice, we use train-test split (90%:10%), five-fold cross-validation, and examining AUC for ROC curves on the 10% held set (as shown in Table 2). Evidently, the LGB model already performs very well with the default parameters only.

*Table 2 Initial Comparison of ML Classification Techniques Using DT, RF, KNN, MLP, GBDT implemented in scikit-Learn Package (SK-GB), GBDT in XGBoost (XGB), and GBDT in LightGBM (LGB).*

| ROC-AUC | n-Seeb | p-Seeb | n-PF | p-PF |
|---|---|---|---|---|
| **DT** | 0.80 | 0.83 | 0.64 | 0.65 |
| **RF** | 0.92 | 0.94 | 0.75 | 0.76 |
| **KNN** | 0.91 | 0.92 | 0.74 | 0.74 |
| **MLP** | 0.73 | 0.95 | 0.74 | 0.76 |
| **SK-GB** | 0.92 | 0.94 | 0.75 | 0.76 |
| **XGB** | 0.92 | 0.95 | 0.75 | 0.76 |
| **LGB** | 0.94 | 0.96 | 0.75 | 0.80 |

We further tune LGB hyperparameters such as the number of estimators, the number of leaves, and the learning rate using a five-fold cross-validation grid search. Using the best model of grid search, we predict the ROC of the 10% held set (shown in Figure 5). We achieve high accuracy for most of the models, with the model for p-type Seebeck being the best one as it corresponds to the highest value of 0.96 as shown in Fig. 5b. We obtain at least 0.8 AUC for most of the ML models, signifying high prediction accuracies.

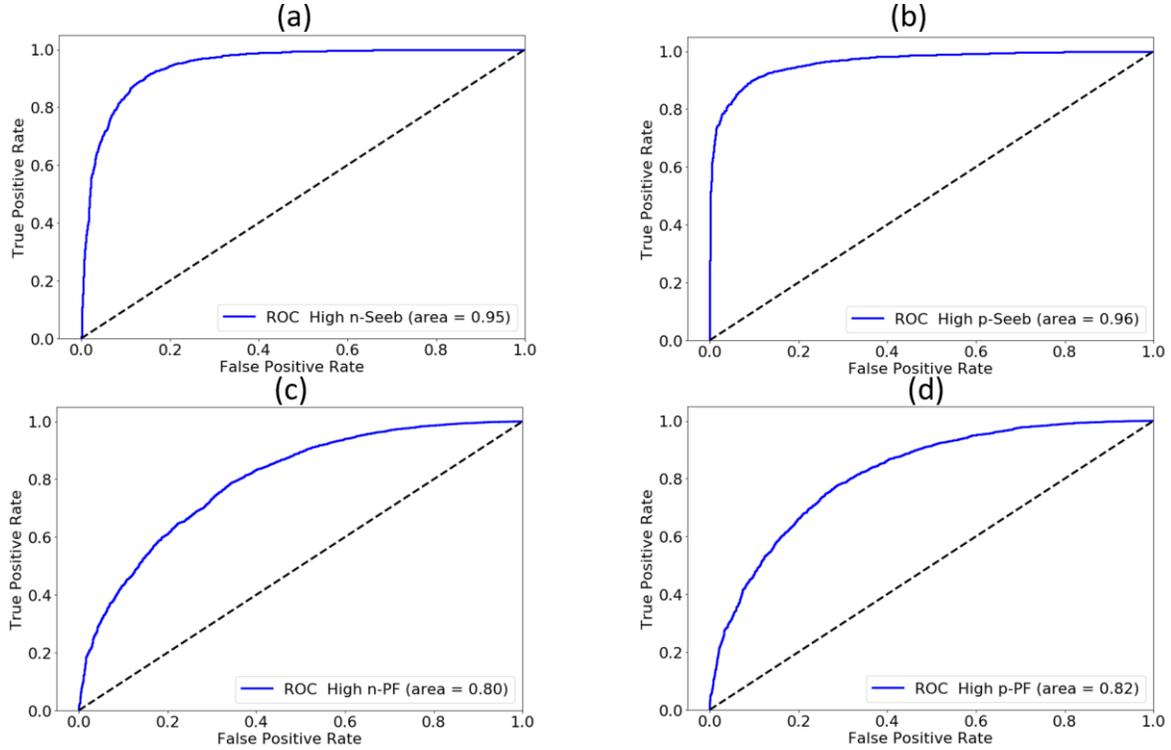

*Fig. 5 Classification receiver operation characteristic curves (ROC) for high Seebeck-coefficient and high power-factor materials. The dotted line shows the random guessing line with an AUC 0.5.*

In addition to the thresholds specified above for classifying high/low performance materials (100 µV/K for |S|, 1000 µW/(mK)$^2$ for PF), we also train models with more stringent thresholds as shown in Table. 3. We find similar ROC AUC for several different thresholds, allowing for more or less sensitive models to be chosen. We can apply these models to arbitrary materials to quickly pre-screen efficient thermoelectric materials, which would become the subject of the next set of DFT calculations. We followed a similar procedure to identify semiconducting 2D materials in our previous work[43] and successfully discovered several 2D materials. Clearly, the screening process can be much accelerated using machine learning models as a first step. All these models are

provided on the JARVIS-ML website (https://www.ctcms.nist.gov/jarvisml/) for predicting thermoelectric performance of new compounds.

*Table 3 Comparison of ROC AUC with varying thresholds used to classify high/low performance materials.*

| Threshold | Model | ROC AUC |
|---|---|---|
| -100 µV/K | p-Seeb | 0.96 |
| -200 µV/K | p-Seeb | 0.96 |
| -300 µV/K | p-Seeb | 0.95 |
| 1000 µW/(mK)$^2$ | p-PF | 0.82 |
| 2000 µW/(mK)$^2$ | p-PF | 0.83 |
| 3000 µW/(mK)$^2$ | p-PF | 0.84 |

Furthermore, the GBDT algorithm allows us to get the feature importance information for each of the 1557 descriptors. We group the features in respective classes[43] such as chemical descriptors (Chem), radial distribution function (RDF), nearest neighbor (NN), angular distribution upto first (ADF-1$^{st}$) and second neighbor (ADF-2$^{nd}$), dihedral distribution function upto first neighbor (DDF), charge descriptor (Chg), cell-size related descriptors (Cell) as shown in Fig. 6. We find that chemical, radial distribution, angle-distribution up to first neighbors, and dihedral angle distribution are almost equally important for achieving a high accuracy model. This is in contrast with the formation energy model[43] in our previous work, where we found that chemistry was the most dominant feature. This might explain why there is no clear trend visible in the periodic table discussed above (see Fig. 3). Out of all the 1557 descriptors, some of the most important descriptors for all the models are: cell-size related descriptors-density and logarithm of volume of the cell, radial distribution peak at 7.5 Å, 9.4 Å and 9.5 Å, first-neighbor based angular distribution peak at 178 degree, mean of product of polarizability and atomic mass, ratio of atomic radii and molar volume and refractive index of individual constituent elements[43].

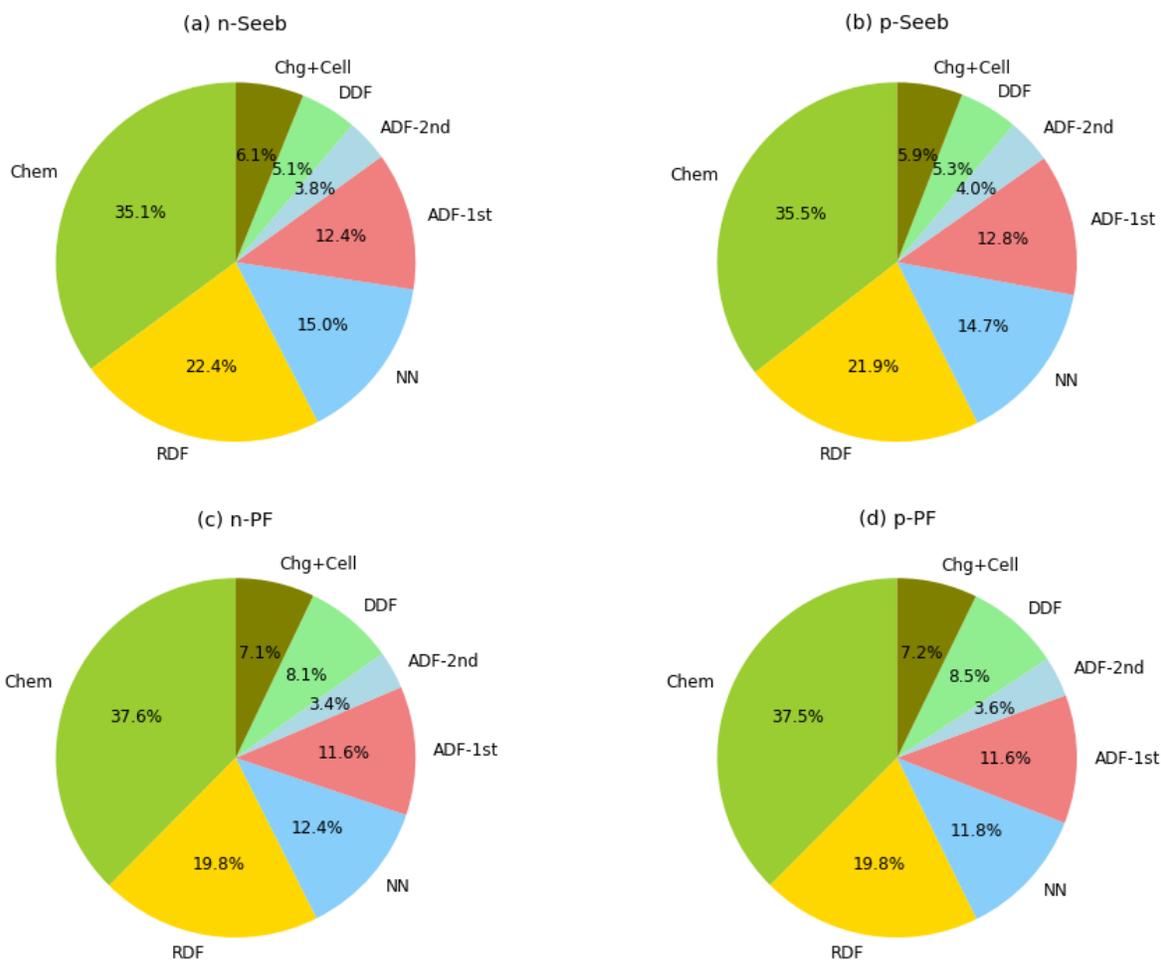

*Fig. 6 Feature importance distribution plot for the classification models.*

**IV CONCLUSIONS**

In summary, we use semiclassical transport methods based on density functional theory calculations to evaluate the thermoelectric properties of both bulk (3D) and monolayer (2D) materials. In addition to identifying interesting candidate materials, we also show chemical, crystallographic, compositional and dimensionality trends for the whole dataset. We screen 2D materials and evaluate trends between the thermoelectric performance of bulk and monolayer geometries. We identify several compositional classes with high thermoelectric performance. We predict ultra-low lattice thermal conductivity in the ZrBrN class of materials. Although the constant-relaxation time approximation is a crude approximation, it allows the generation of large-scale database for initial screening of thermoelectric materials. Finally, we train machine learning models to accelerate the future screening processes. We believe that our data and tools for evaluating and predicting thermoelectric performance will greatly enhance the discovery and characterization of thermoelectric materials.

**V SUPPLEMENTARY MATERIAL**

See the supplementary material for the comparison of theoretical and experimental data as well as the dataset generated in the present work.